# Orbital-angular-momentum mode-group multiplexed transmission over a graded-index ring-core fiber based on receive diversity and maximal ratio combining

**JunweiZhang**[1,†]**, Guoxuan Zhu**[1,†]**, Jie Liu**[1,*]**, Xiong Wu**[1]**, Jiangbo Zhu**[2]**, Cheng Du**[3]**, Wenyong Luo**[3]**, and Siyuan Yu**[1,2]

[1]*State Key Laboratory of Optoelectronic Materials and Technologies, School of Electronics and Information Technology, Sun Yat-Sen University, Guangzhou 510006, China*
[2]*Photonics Group, Merchant Venturers School of Engineering, University of Bristol, Bristol BS8 1UB, UK*
[3]*Fiberhome Telecommunication Technologies Co. Ltd,Wuhan, 430074, China*
[†]*These authors contributed equally to this work.*

**Abstract**: An orbital-angular-momentum (OAM) mode-group multiplexing (MGM) scheme based on a graded-index ring-core fiber (GIRCF) is proposed, in which a single-input two-output (or receive diversity) architecture is designed for each MG channel and simple digital signal processing (DSP) is utilized to adaptively resist the mode partition noise resulting from random intra-group mode crosstalk. There is no need of complex multiple-input multiple-output (MIMO) equalization in this scheme. Furthermore, the signal-to-noise ratio (SNR) of the received signals can be improved if a simple maximal ratio combining (MRC) technique is employed on the receiver side to efficiently take advantage of the diversity gain of receiver. Intensity-modulated direct-detection (IM-DD) systems transmitting three OAM mode groups with total 100-Gb/s discrete multi-tone (DMT) signals over a 1-km GIRCF and two OAM mode groups with total 40-Gb/s DMT signals over an 18-km GIRCF are experimentally demonstrated, respectively, to confirm the feasibility of our proposed OAM-MGM scheme.

## References and links

## 1. Introduction

Space-division multiplexing (SDM) in optical fiber has recently been intensively investigated, aiming for solving the current single mode fiber (SMF) capacity crunch by utilizing the spatial or mode domain of light [1]. Among various SDM schemes, mode-division multiplexing (MDM) techniques based on multi-mode fibers (MMFs) or few-mode fibers (FMFs) can increase the number of transmission channels within a limited aperture and thus increase the capacity density of a single fiber core [2–4]. In addition, the design of amplifiers, switches and other inline components in MDM schemes can be highly compact, which makes the scaling of optical networks more cost effective and energy efficient [4, 5]. The main limitations of MMF-based MDM systems are the crosstalk and distortion resulting from mode coupling and modal dispersion during fiber transmission. In long-haul MDM systems, crosstalk between all mode pairs is non-negligible [6]. As a result, adaptive full-size multiple-input multiple-output (MIMO) equalization is required, in which case the fiber in the strong-mode-coupling regime is even more desirable to decrease the differential group delay (DGD) and thus reduce the complexity of MIMO processing in these systems [7]. However, on the other hand, in short-reach applications (e.g. intra-data-center network, local area network, access network, etc.), intensity-modulated direct-detection (IM-DD) schemes without MIMO processing are preferred, considering the system cost and power consumption [8]. Weakly coupled MDM scheme can be considered as one of the most promising solutions to increase the capacity of short-reach transmission systems, since modal crosstalk and dispersion can be neglected for short-reach transmission and the need of coherent optical detection and MIMO processing can thus be eliminated [9, 10]. However, reducing mode coupling among all the fiber modes, especially the (quasi-)degenerate modes, over > 2-km fiber distance still remains a challenge in these schemes.

Given the increasing mode coupling in MMFs over distance, Mode-group multiplexing (MGM) [11] emerges as an alternative for MIMO-free MMF transmission, in which (quasi-) degenerate modes within each mode group (MG) are regarded as one data channel. The weak-coupling strategy can be employed between different MGs to eliminate the need of coherent detection and MIMO processing at the receivers. Several MGM schemes have been implemented based on conventional MMFs, in which all the de-multiplexed intra-group modes should be detected simultaneously at the receiver to avoid the mode partition noise resulting from the random intra-group mode crosstalk [12–14]. However, as the number of intra-group modes of the MMFs linearly increases with the MG order [as shown in Fig. 1(a)], reception of high-order MGs will become increasingly complex, which limits scalability of the MMF-based MGM systems.

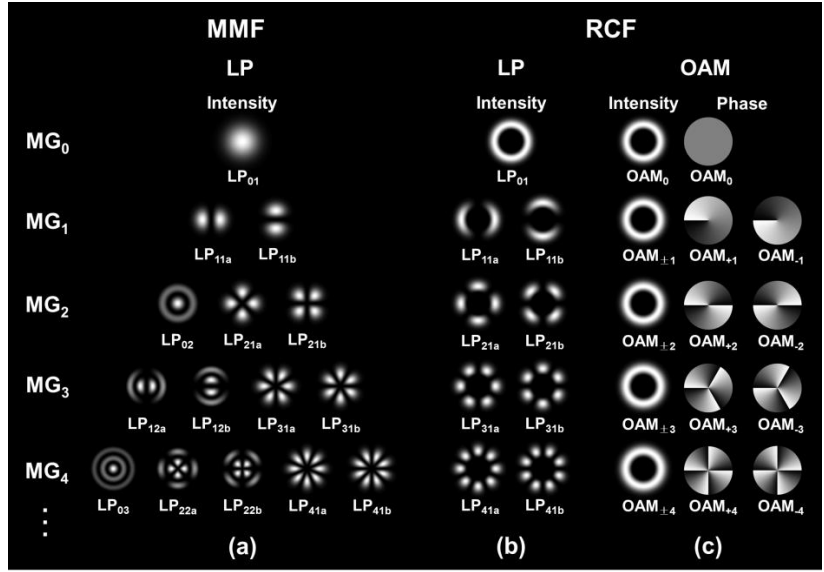

Fig. 1. Diagrams of (a) intensity of the linearly polarized (LP) modes in each mode group (MG) of the multimode fiber (MMF); (b) intensity of the LP modes in each MG of the ring-core fiber (RCF); (c) intensity and phase of the OAM modes in each MG of the RCF. Here note that each fiber mode shown in this figure can be further subdivided into two modes with orthogonal polarizations.

Ring-core-fiber (RCF) based MGM systems have been reported recently [15–17]. With the ring-core profile allowing for only single-radial-order modes, a fixed number of degenerate modes are supported by the RCFs for each high order MG [as shown in Fig. 1(b) and (c)], which decreases the complexity of the high-order MG detection. In addition, the coupling strength between adjacent MGs of the RCFs decreases significantly with the increasing azimuthal mode order [18]. These characteristics provide RCFs with a higher scalability in the optical mode space. Besides, both the linearly polarized (LP) and orbital angular momentum (OAM) mode basis can be utilized for MGM in RCFs. MG

(de)multiplexing of the RCF-based MGM schemes that have been reported is mainly based on weighted composite phase masks, in which the relative amplitude and phase of all intra-group modes should be measured at the receiving end [15–17]. However, as the relative amplitude and phase of the intra-group modes change randomly during fiber transmission, adaptive measurement or evaluation of the amplitude and phase distribution of degenerate modes in optical domain may be quite difficult in practical implementations.

Therefore, in this paper, an OAM-MGM system over a graded-index ring-core fiber (GIRCF) base on a receive-diversity scheme is proposed. In the receive-diversity scheme, a single-input two-output architecture is designed for each MG channel and simple digital signal processing (DSP) is utilized to adaptively resist the mode partition noise induce by random intra-group mode crosstalk. There is no need of adaptive MIMO equalization in this scheme. Furthermore, a simple maximal ratio combining (MRC) technique can be utilized at the receiver of this scheme to efficiently take advantage of the diversity gain of receiver and improve the signal-to-noise ratio (SNR) of the received signals. In order to prove the feasibility of the proposed OAM-MGM scheme, intensity-modulated direct-detection (IM-DD) systems transmitting three OAM mode groups with total 100-Gb/s discrete multi-tone (DMT) signals over a 1-km GIRCF and two OAM mode groups with total 40-Gb/s DMT signals over an 18-km GIRCF are experimentally demonstrated, respectively. The measured results show that by using the receive-diversity architecture and MRC technique, there will be an average of ~3 dB improvement of sensitivity at the BER of $3.8\times10^{-3}$ for both of the OAM-MGM system over 1-km and 18-km GIRCF, compared with that in the OAM-MGM system based on single-PD detection. In addition, when the receive-diversity architecture is utilized, MRC-based system performance is superior to that of the system with equal ratio combining (ERC), especially in the case that there is a great difference of BER performance between the two received branches.

## 2. Proposed OAM-MGM scheme

The block diagram of the proposed OAM MGM is shown in Fig. 2. For each MG, optical light at a fixed wavelength is intensity modulated by an electrical signal to generate the optical signal at the transmitter. Considering the linear or quasi-linear relationship between the optical light intensity and the electrical signal, the electric field of the intensity modulated optical signal for the $i_{th}$ MG can be expressed as:

$$E_i(t) = A_i \sqrt{\alpha_i[V_{i0} + V_i(t)]} e^{j(\omega_0 t + \varphi_i)} \tag{1}$$

where $A_i$, $\omega_0$ and $\varphi_i$ denote the amplitude, frequency and phase of the optical carrier, respectively, while $\alpha_i$ is the ratio between optical power and electrical power before and after electro-optic conversion at the transmitter of $i_{th}$ MG and here can be considered as a constant. $V_{i0}$ and $V_i(t)$ refer to the DC and AC components of the electrical signal carried by the $i_{th}$ MG channel, respectively. The generated optical signals of all MGs are launched to their respective SMF input ports of the OAM multiplexer. Here note that, for each OAM MG that includes four degenerate OAM modes $<\pm l, \pm s>$ ($\pm s$ being the left- or right-hand circular polarizations and $\pm l$ being the azimuthal mode order) [19], only one OAM mode is excited at the transmitter, as shown in Fig. 2. Then the optical signals are OAM mode converted and multiplexed with an OAM multiplexer (OAM mux, e.g. the OAM mode sorter [20]), and finally coupled to the GIRCF. Since the GIRCF here is designed with large inter-group effective index differences but very small effective index differences between intra-group modes [19], random intra-group mode crosstalk resulting from strong modal coupling is inevitable during fiber transmission, while there is only low coupling between different MGs. As a result, all the four intra-group modes should be simultaneously detected at the receiver to avoid the mode partition noise, while small power loss due to weak coupling between MGs, which is proportional to the fiber length and does not vary with time randomly [21], can be neglected. The electric field of four degenerate modes in the $i_{th}$ MG after a certain–distance GIRCF transmission can be expressed as:

$$E_{\text{OAM}_{l,m}}(t) = A'_{l,m}\sqrt{\alpha_i[V_{i0} + V_i(t)]}e^{j(\omega_0 t + \varphi'_{l,m})}e^{jl\theta} \quad (2)$$

where $l$ equals to $\pm i$ (the azimuthal mode order) and $m$ equals to $\pm s$ (the left- or right-hand circular polarizations). $A'_{l,m}$ and $\varphi'_{l,m}$ denote the amplitude and phase of the OAM$_{l,m}$ mode, respectively, whose values randomly change in time because of random crosstalk between intra-group modes. $\theta$ is the azimuthal angle. As power loss of the $i_{th}$ MG resulting from fiber loss and inter-group crosstalk is related to the fiber length, given a certain fiber structure and length, the total optical power of all the four modes of the $i_{th}$ MG:

$$\begin{aligned}P_{\text{MG}_i} &= \sum_{l=\pm i}\sum_{m=\pm s}\left|E_{\text{OAM}_{l,m}}(t)\right|^2 \\ &= \alpha_i[V_{i0} + V_i(t)]\sum_{l=\pm i}\sum_{m=\pm s}\left|A'_{l,m}\right|^2 = C\alpha_i[V_{i0} + V_i(t)]\end{aligned} \quad (3)$$

where $C$ is a constant. As a result, the total optical power of the $i_{th}$ MG is proportional to the electrical signal, which theoretically proves that the mode partition noise due to random modal crosstalk can be eliminated when all the four intra-group modes are simultaneously power detected.

After mode converted and demultiplexed by the OAM demultiplexer (OAM demux), two polarization multiplexed fundamental modes at each SMF output port of the OAM demux, which are converted from the two OAM modes with the same azimuthal mode order but orthogonal polarizations, respectively, are detected by one photo detector (PD). Considering the different power response of the two received branches, the detected photo current after square-law detection of the $i_{th}$ MG is

$$\begin{aligned}I_{\text{MG}_i} &= \mu_{+i}\sum_{m=\pm s}\left|S_{\text{OAM}_{+i,m}}(t)\right|^2 + \mu_{-i}\sum_{m=\pm s}\left|S_{\text{OAM}_{-i,m}}(t)\right|^2 \\ &= \alpha_i[E_{i0} + E_i(t)](\mu_{+i}\sum_{m=\pm s}\left|A'_{+i,m}\right|^2 + \mu_{-i}\sum_{m=\pm s}\left|A'_{-i,m}\right|^2)\end{aligned} \quad (4)$$

where $\mu_{+i}$ and $\mu_{-i}$ are responsivity of the two received branches, respectively, considering power response of both the optical transmission paths and PDs. Here note that optical beams from SMF$_{\pm i}$ output ports of the OAM demux are not directly added together in optical domain and detected by a same PD in order to avoid optical interferences between non-orthogonal modes. From Eq. (4), one can see that if signals from the two received branches of the $i_{th}$ MG are directly combined, the detected electrical signal is no longer proportional to the optical power, due to the different power response of the two received branches. In addition to the power response difference, there are other two kinds of channel impairments to the two-branch signal reception of the $i_{th}$ MG: 1) the relative delay between the two received branches induced by the differential modal delay, which will desynchronize the received signals from the two different branches; 2) randomly varied signal-to-noise ratio (SNR) of the two received signals caused by the random power coupling between the $\pm i$ OAM modes, which will deteriorate SNR performance of the detected signal if electrical signals of the two received branches are directly added together with equal weight.

In order to deal with these problems, low-complexity digital signal processing (DSP) should be employed in the electrical domain, as shown in Fig. 2. For each received branch of the $i_{th}$ MG, electrical signal from the PD is first digitalized by an analog-to-digital convertor (ADC) and then launched to the DSP module. In the DSP module, after resampled and symbol synchronized, signal of each received branch of the $i_{th}$ MG is equalized using either time domain equalization (TDE) or frequency domain equalization (FDE) algorithms [8] to compensate the distortions resulting from the differential modal delay and chromatic dispersion. Here note that the TDE or FDE is implemented for the single-channel equalization rather than MIMO equalization, while the former has a much lower calculation complexity [8, 22]. In order to maximize the SNR of the received signal regardless of the power fluctuation and the responsivity difference of the two received branches, the two equalized and demodulated signals of the same MG are combined by using a simple MRC algorithm [23]. The MRC output of the $n$th symbol for each data frame is

$$z(n) = \sum_{l=\pm i}(y_l(n)\times \text{SNR}_l)/\sum_{l=\pm i}\text{SNR}_l \qquad (4)$$

where $z(n)$ is the combined signal of the $n^{th}$ symbol used for final symbol decision and demapping to obtain the bit data of the $i_{th}$ MG, $y_l(n)$ with $l = \pm i$ are two equalized signals of the $n^{th}$ symbol, and $\text{SNR}_l$ with $l = \pm i$ are the probed SNRs of the two equalized signals obtained by training symbol and assumed to be quasi-time-invariant within one signal frame. It should be noted that this MRC algorithm can be executed to combine two dependent signals, either in the time domain for pulse-amplitude modulation (PAM) and carrier-less amplitude and phase (CAP) modulation, or in the frequency domain for DMT modulation before final symbol decision at the receiver. It is noted that in the proposed OAM-MGM scheme only two received branches rather than four branches for the signal reception of each high order MG in order to decrease the receiver complexity. Although equal-weight combination of signals carried by the two polarization multiplexed modes in each received branch might deteriorate the SNR performance of the received signals, there should be a trade-off between the signal performance and the system complexity in practical implementations.

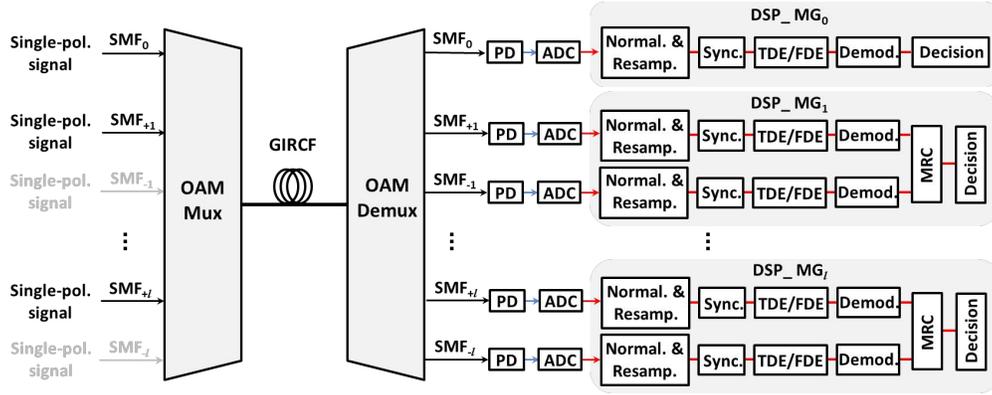

Fig. 2. Block diagram of the proposed OAM mode-group (de)multiplexing scheme. SMF$_n$: the single mode fiber input/output port for the $n_{th}$ OAM modes; MG$_n$: the $n^{th}$ OAM mode group including OAM modes $<\pm l, \pm s>$; Normal.: Normalization; Resamp.: resampling; demod.: demodulation.

## 3. Fiber design and characterization

For proof-of-concept demonstration of the proposed OAM-MGM scheme, a GIRCF supporting mode group order up to $|l| = 4$ is designed and fabricated. As shown in Fig. 3(a), the GIRCF has a parabolic index profile over the ring-core width to: 1) provide large inter-MG differential effective refractive indices ($\Delta n_{eff}$) and low intra-MG $\Delta n_{eff}$ thereby de-coupling MGs and decreasing differential modal delay between intra-MG modes[see Fig. 3(b) and (c)], 2) soften the radial index gradient thus making the fiber less susceptible to perturbations such as micro-bending [24], and 3) eliminate the spin-orbit-coupling-induced mode purity impairment [25] by removing the step refractive index (RI) interface. It can be seen from Fig. 3(a) that the maximum material RI difference $\Delta n$ is around 0.027, the ring-core radius is around 6.6 μm and the ring-core width is around 3.3 μm. The calculated effective RI of all guided OAM modes and the calculated/measured DGD at the wavelength of 1550 nm are illustrated in Fig. 3(b) and (c), respectively. The $\Delta n_{eff}$ and DGD between adjacent MGs increase with the topological order of MGs, and high order MGs ($|l| > 1$), which will be selected for the demonstration below, promise a higher resistance to inter-MG crosstalk. In addition, the very low intra-MG DGD for all MGs (from both calculation and measurements) indicates a low memory-size requirement on channel equalization.

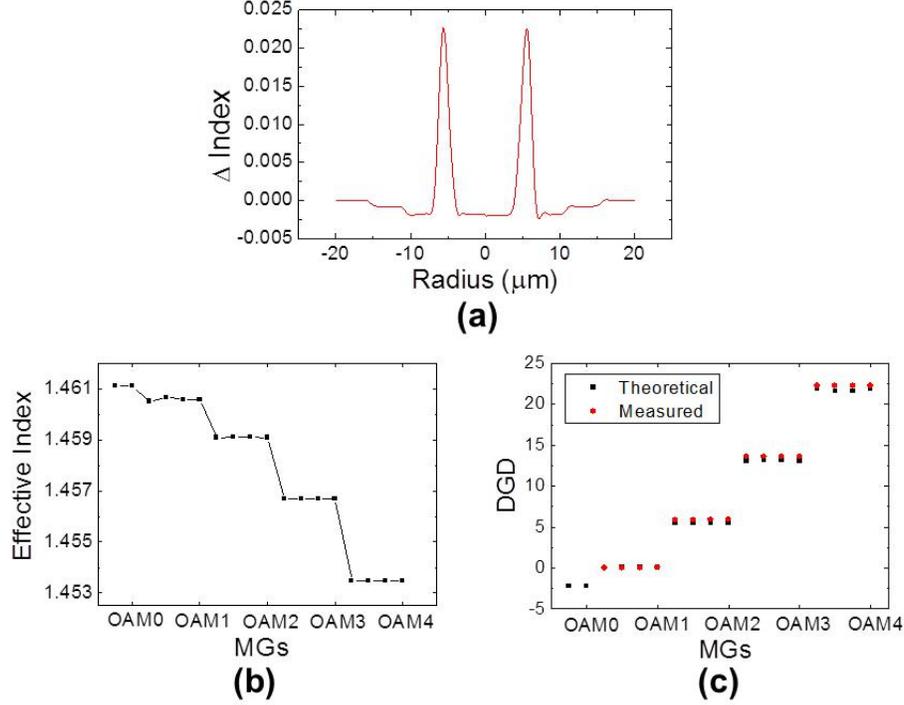

Fig. 3. (a) The refractive index profile of GIRCF; (b) effective indices of MGs in the GIRCF at the wavelength of 1550nm; (c) calculated and measured DGD at the wavelength of 1550 nm.

The GIRCF is fabricated by using a conventional plasma enhanced chemical vapor deposition (PECVD) process. The measured average propagation attenuation of this fiber is 0.75 dB/km at 1550 nm. This high loss might be caused by the imperfections on two graded index interfaces of the ring core. The sub-system consisting of the GIRCF and the OAM (de)multiplexer (Mux/Demux) devices (i.e., the part in the central dashed box in Fig. 4) is also built to characterize the crosstalk amongst higher order MGs experimentally. The measured results of the 1-km and 18.4-km GIRCF based systems are shown in Table I and II, respectively. It can be deduced consequently that the pure in-fiber mode-coupling induced crosstalk between MGs $|l|$ = 3 & 4 over a 17.4-km GIRCF is ~ -9.65 dB in average, while that between MGs $|l|$ = 2 & 3 is ~ -7.58 dB. Considering the relatively larger inter-MG crosstalk between MGs $|l|$ = 2 & 3 in the 18.4-km GIRCF based system, the two high-order MGs $|l|$ = 3 and 4 are selected for the OAM-MGM transmission, while three MGs $|l|$ = 2, 3 and 4 in the 1-km GIRCF based system are utilized for data transmission demonstration (details will be discussed in next section).

**Table I. The static inter-MG crosstalk (in dB) of entire optical system with 1-km GIRCF.**

| 1-km GIRCF | | Destination MG | | | |
|---|---|---|---|---|---|
| | | $|l|=1$ | $|l|=2$ | $|l|=3$ | $|l|=4$ |
| Source MG | $|l|=1$ | 0 | -4.43 | -15.03 | -18.57 |
| | $|l|=2$ | -5.09 | 0 | -11.26 | -18.04 |
| | $|l|=3$ | -17.36 | -11.94 | 0 | -14.05 |
| | $|l|=4$ | -20.99 | -17.29 | -13.8 | 0 |

**Table II. The static inter-MG crosstalk (in dB) of entire optical system with 18.4-km GIRCF.**

| 18.4-km GIRCF | | Destination MG | | |
|---|---|---|---|---|
| | | $|l|=2$ | $|l|=3$ | $|l|=4$ |
| Source MG | $|l|=2$ | 0 | -5.19 | -7.66 |
| | $|l|=3$ | -7.33 | 0 | -8.68 |
| | $|l|=4$ | -8.39 | -7.9 | 0 |

### 4. Data transmission demonstration

An IM-DD DMT transmission system is built for the demonstration of the proposed OAM-MGM scheme. The experimental setup is shown in Fig. 4. At the transmitter, the input data bit sequences are firstly mapped into quadrature amplitude modulation (QAM) symbols. After serial-to-parallel (P/S) conversion, the QAM symbols are converted to time domain by 2536-point inverse FFT (IFFT). In order to generate a real-valued DMT signal, the 9th – 264th subcarriers and the 2029th – 1774th subcarriers are used to carry the effective payloads and their complex conjugates, respectively. Then 48-point cyclic prefix (CP) padding, parallel-to-serial (P/S) conversion and hard clipping are performed. For each DMT frame, 11 training symbols are inserted at the beginning of the data frame, which consist of 1 symbol for timing synchronization and 10 symbols for channel estimation and SNR probing. By using an arbitrary waveform generator (AWG) operating at a sample rate of 60-Gsa/s, the electrical DMT signal with an effective bandwidth of 10-GHz is generated. After amplification, the electrical signal is utilized to modulate optical light at wavelength of 1550.92 nm though a Mach-Zehnder modulator (MZM) to generate double-sideband optical signal. Then the generated optical signal is split into three branches, each of which is amplified by an erbium doped fiber amplifier (EDFA) and delayed by a single mode fiber (SMF) with large relative length from other two branches for data pattern decorrelation.

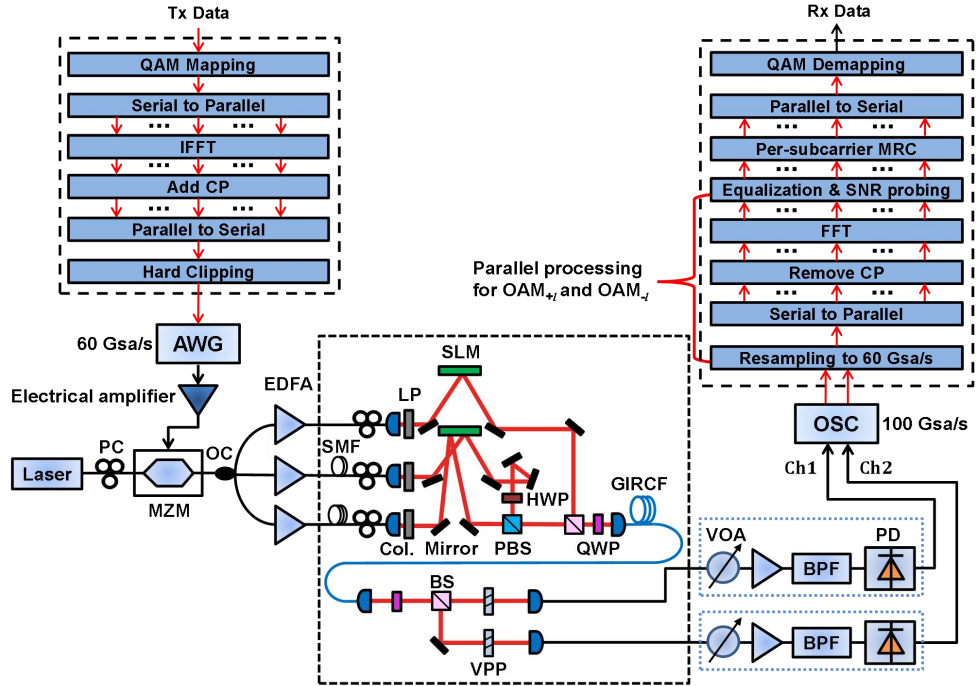

Fig. 4. Experimental setup. AWG: arbitrary waveform generator; PC: polarization controller; MZM: Mach-Zehnder modulator; OC: optical coupler; SMF: single-mode fiber; LP: linear polarizer; SLM: spatial light modulator; PBS: polarizing beam splitter; HWP: half-wave plate; QWP: quarter-wave plate; Col.: collimator; BS: beam splitter; VPP: vortex phase plate; VOA: variable optical attenuator; EDFA: erbium doped fiber amplifier; BPF: band pass filter; PD: photo detector; OSC: oscilloscope.

After being collimated and linearly polarized, the Gaussian beams from the three SMF branches are converted to OAM modes of $l = +2, +3$ and $+4$, respectively, by using the phase-only spatial light modulators (SLMs). Here note that due to the device limitation, conversions of OAM modes of $l = +3$ and $+4$ are realized by spatially sharing one SLM. By employing a polarization beam splitter (PBS), the two OAM beams of $l = +3$ and $+4$ possessing orthogonal linear-polarizations are combined with low loss. Then the two multiplexed OAM beams are combined with the OAM beam of $l = +2$ by a beam splitter. The generated three coaxial OAM beams are converted into circular polarization states through a quarter-wave plate (QWP) before coupled into the GIRCF, since the eigen OAM state in ring-core fiber is circular polarized. After GIRCF transmission, all output modes from the fiber are converted into linear polarizations and split into two branches. In each branch, the OAM beams are converted to the Gaussian beams by a vortex phase plate (VPP) and then collimated to the SMF pigtail for photo-electric detection. Here it should be noted that the two received branches are used for detection of $\pm l$ OAM modes, respectively, for the reception of the $l_{th}$ MG.

In each received branch, the optical signal is detected by an optically pre-amplified receiver, which consists of a variable optical attenuator (VOA), an EDFA followed by an optical band-pass filter and a PD. Then the detected electrical signals are digitized and stored by a real time oscilloscope (OSC) with a sampling rate of 100 Gsa/s and finally processed by off-line DSP including resampling, timing synchronization, serial-to-parallel (S/P) conversion, FFT, one-tap channel equalization, per-subcarrier MRC [23], demapping and error counting are used to process the data.

We performed two sets of MGM transmission over the GIRCFs: transmission of three OAM MGs over a 1-km GIRCF and two OAM MGs over an 18-km GIRCF. The back-to-back configuration has also been implemented with a short length of GIRCF (~2-m) between the OAM MUX and DEMUX for system performance comparison, and the BER for each MG are measured and evaluated individually. Fig. 5(a)-(c) show the measured BER results as a function of the received optical power (ROP) of the first MGM transmission system, in which three adjacent OAM MGs of $|l|$ = 2, 3 and 4 carrying 10-Gbaud DMT signals are transmitted over a 1-km GIRCF. Here note that the modulation format of the DMT signals carried by OAM MG of $|l|$=2 is quadrature phase shift keying (QPSK), while that of the signals carried by other two MGs is 16QAM. The following observations could be made from Fig. 5(a)-(c): 1) Compared with that in the back-to-back case, power penalty of the MG $|l|$ = 2, 3 and 4 for single-MG transmission by using MRC technique based on two received branches ($|l|$ = 2, 3 or 4 only, w/ MRC) over the 1-km GIRCF are 11.5 dB, 1.1 dB and 2.6 dB, respectively, at BER of $3.8\times10^{-3}$. The MG $|l|$ = 2 suffers a much higher power penalty due to relatively stronger inter-MG coupling between MG $|l|$ = 1 and 2 (see Table I). 2) Compared with the single-MG transmission case, power penalty of the MG $|l|$ = 2, 3 and 4 at BER of $3.8\times10^{-3}$ for three-MG transmission by using MRC technique ($|l|$ = 2, 3 or 4 w/ MRC) over the 1-km GIRCF are 1 dB, 4.5 dB and 0.7 dB, respectively, which implies that the MG $|l|$ = 3 suffer more crosstalk from the other two MGs, compared with those of the MG $|l|$ = 2 and 3. 3) Due to the receiver-diversity gain and SNR improvement from MRC, there is a ~2 dB and ~5 dB power budget improvement at BER of $3.8\times10^{-3}$ for MG $|l|$ = 2 and 4 by using the MRC technique, respectively, compared with the case based on single-PD detection ($|l|$ = $i$, receive $l$ = +$i$ or $l$ = -$i$, $i$ = 2 and 4). As for the MG $|l|$ = 3, when the receive diversity architecture (two received branches) and MRC technique are utilized, BER of $3.8\times10^{-3}$ for three-MG transmission over the 1-km GIRCF can be realized at ROP of ~-14 dBm, while the BER in the case of single-PD detection fail to achieve the 7% hard-decision forward error correction (FEC) limit of $3.8\times10^{-3}$ at ROP less than -10 dBm. 4) The BER performance of system based on two received branches and equal ratio combining (ERC) technique ($|l|$ = 2, 3 or 4 w/ ERC) is also evaluated for comparison. It can be seen that the there is a similar BER performance for the cases w/ MRC and w/ ERC (the case w/ MRC is slightly better than that of w/ ERC) when the BER performance of $l$ = +$i$ and –$i$ in the case of single-PD detection are approximately same [see Fig. 5(b) and (c)]. However, when the BER performance of $l$ = +$i$ and –$i$ have a great difference, there will be an improvement of BER performance for the case w/ MRC compared with that of w/ ERC [see Fig. 5(a)]. Here note that the BER curves in the case of single-PD detection have larger fluctuations, which can be ascribed to severe power fluctuation induced by strong coupling among intra-MG modes. It can be deduced from the results that the system with MRC has much more stable performance compared with that of the system with ERC, considering the random power crosstalk among intra-MG modes. The received constellation diagrams for MG $|l|$ = 3 at ROP of -10.7 dBm after 1-km RCF transmission are shown in Fig. 5(d). Here three receiving schemes are considered, which are single-PD-detection scheme, two-PD-detection scheme w/ ERC and two-PD-detection scheme w/ MRC. It can be seen from the results that the scheme w/ MRC has the best performance among the three cases.

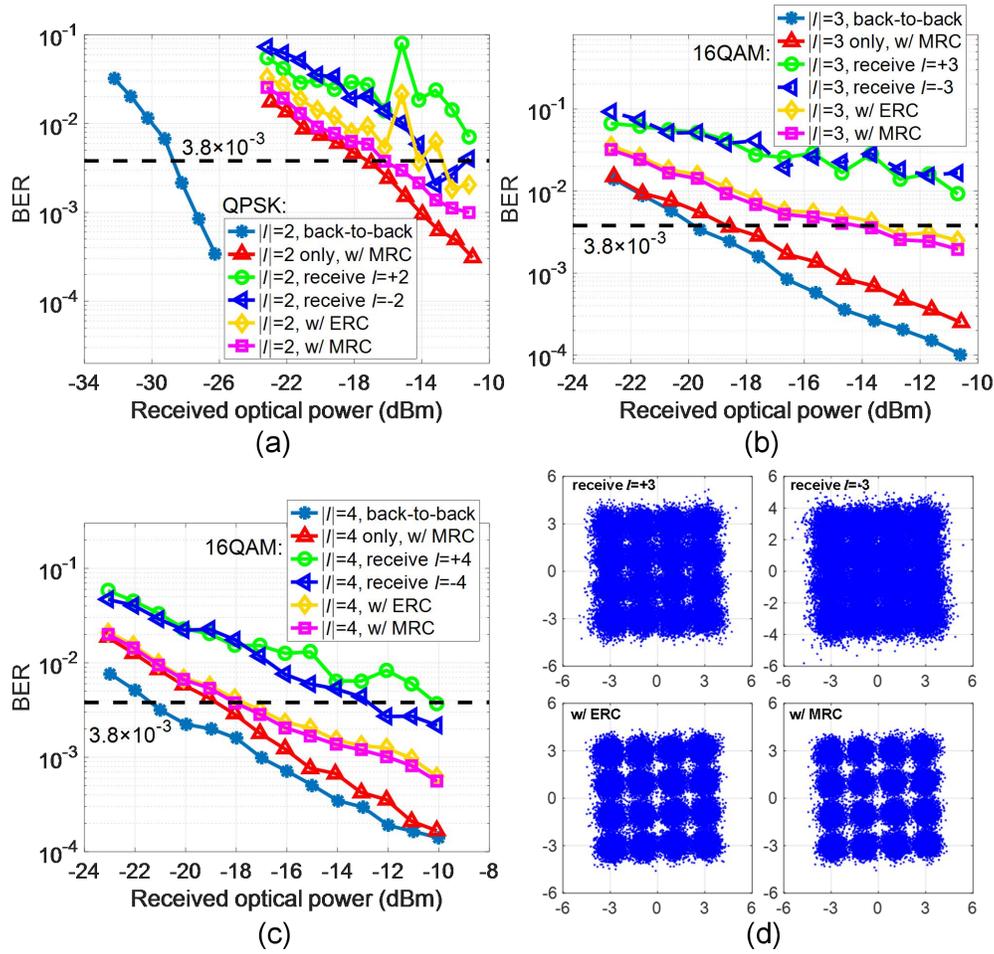

Fig. 5. Measured BER versus ROP for (a) MG $|l| = 2$, (b) MG $|l| = 3$, and (c) MG $|l| = 4$ in the OAM-MGM transmission system over a 1-km GIRCF; (d) received constellation diagrams for MG $|l| = 3$ at a ROP of -10.69 dBm after 1-km GIRCF transmission.

Fig. 6(a) and (b) present the measured BER results as a function of the ROP for two mode-group multiplexed transmission over an 18-km GIRCF. In this system, two adjacent MGs $|l| = 3$ and 4 are utilized to transmit DMT signals with modulation format of QPSK. It can be seen from the results in Fig. 6(a) and (b) that: 1) Compared with that in the single-MG transmission case, there is ~2 dB and ~0.9 dB power penalty at BER of $3.8 \times 10^{-3}$ for MG $|l| = 3$ and 4, respectively, in the case of two-MG transmission over the 18-km GIRCF. 2) By utilizing the two-PD-detection architecture and MRC technique, the received sensitivity at BER of $3.8 \times 10^{-3}$ for MG $|l| = 3$ and 4 can be improved more than 2 dB and 4 dB, respectively, compared with the single-PD-detection cases. 3) Compared with the two-PD-detection scheme w/ ERC, the scheme with MRC has a better performance for both the MGs $|l| = 3$ and 4 [see Fig. 6(a) and (b)], especially when BER performance of $l = +i$ and $-i$ has a great difference [see Fig. 6(a)].

Fig. 6(c) and (d) show the SNRs and BERs of individual subcarriers of MG $|l| = 3$ used for carrying effective payload at a ROP of -20.9 dBm after 18-km RCF transmission. One can see that the SNR or BER distributions of two single-PD-detection cases are quite different due to the different received powers resulted random power crosstalk between OAM modes of $l = +i$ and $-i$ and different responses of the two received branches. The SNR in the two-PD-detection system w/ MRC outperforms that in the two-PD-detection system, especially when the SNR performance for $l = +i$ and $-i$ have a great difference, which agree with the results shown in Fig. 6(a) and (b). It should be noted that the SNRs of high frequency subcarriers are lower than that of low frequency subcarriers when MRC technique is employed, which results in better BER performance for low frequency subcarriers. In order to achieve a higher capacity, adaptive bit and power loading algorithm, which can flexibly allocate bits and power for each subcarrier in terms of SNR distribution, could be assigned in future study.

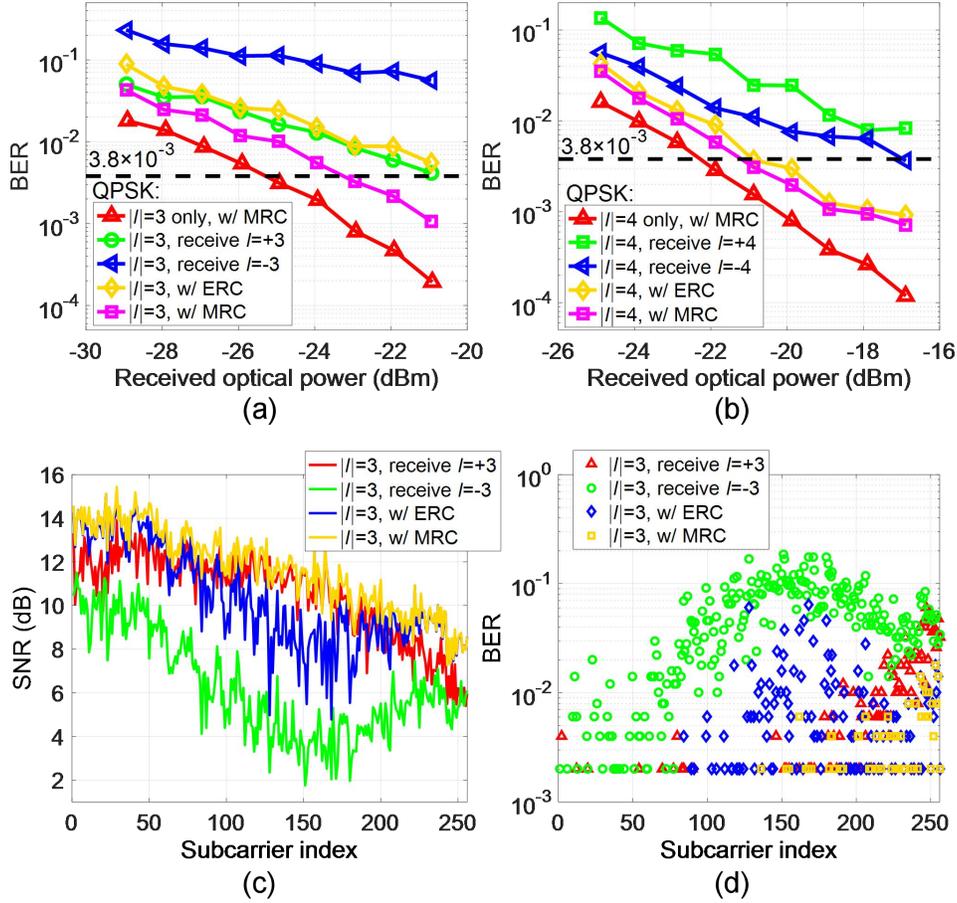

Fig. 6. Measured BER versus ROP for (a) MG $|l| = 3$ and (b) MG $|l| = 4$ in the OAM-MGM transmission system over an 18-km GIRCF; (c) SNRs and (d) BERs of individual subcarriers of MG $|l| = 3$ used for carrying effective payload at a ROP of -20.9 dBm after 18-km RCF transmission.

## 5.  Conclusions

In this paper, we have proposed and experimentally demonstrated an OAM-MGM scheme based on a GIRCF by utilizing simple MRC at the receivers. To resist the mode partition noise resulted from the random intra-group mode crosstalk, a receive-diversity architecture has been designed for each MG channel. Moreover, a simple MRC technique has been employed on the receiver side, in order to improve the SNR of the received signals by making use of the diversity gain of receiver, without performing complex MIMO processing. To confirm the feasibility of our proposed OAM MG multiplexing scheme, IM-DD schemes transmitting three OAM mode groups with total 100-Gb/s DMT signals over a 1-km GIRCF and two OAM mode groups with total 40-Gb/s DMT signals over an 18.4-km GIRCF have been experimentally demonstrated, respectively. The measured results show that the MRC-based system exhibits best BER performance among the three schemes, which are systems with single-PD detection, with ERC-based two-PD detection (receive diversity), as well as with MRC-based two-PD detection.

**Funding**

SYSU is supported by National Basic Research Program of China (973 Program) (2014CB340000), National Natural Science Foundations of China (61490715, 61505266, 61323001, 11690031, 51403244), Guangdong Natural Science Foundation (2014A030310364, 2016A030313289) and Science and Technology Program of Guangzhou (201707020017). UoB is supported by European Union Horizon2020 project ROAM.